\documentclass[conference]{IEEEtran}
\IEEEoverridecommandlockouts
\usepackage{cite}
\usepackage{amsmath,amssymb,amsfonts}
\usepackage{algorithmic}
\usepackage{graphicx}
\usepackage{textcomp}
\usepackage{xcolor}
\usepackage{tabularx} 
\usepackage{makecell}
\usepackage{booktabs} 
\usepackage{cite}
\usepackage{graphicx}
\usepackage{float}
\usepackage{multirow}
\usepackage{color}
\usepackage{rotating}
\usepackage{tabularray}
\usepackage{authblk}
\usepackage{hyperref}
\def\BibTeX{{\rm B\kern-.05em{\sc i\kern-.025em b}\kern-.08em
    T\kern-.1667em\lower.7ex\hbox{E}\kern-.125emX}}

\begin{document}

\title{Ultrasound-QBench: Can LLMs Aid in Quality Assessment of Ultrasound Imaging?}

\author{
    Hongyi Miao$^1$, Junjia Liu$^2$\textdagger, Yankun Cao$^1$, Yingjie Zhou$^2$, Yanwei Jiang$^2$, Zhi Liu$^1$\textdagger, Guangtao Zhai$^2$\\
    $^1$Shandong University, $^2$Shanghai Jiao Tong University\\
    \textdagger Corresponding author
}

\maketitle

\begin{abstract}
With the dramatic upsurge in the volume of ultrasound examinations, low-quality ultrasound imaging has gradually increased due to variations in operator proficiency and imaging circumstances, imposing a severe burden on diagnosis accuracy and even entailing the risk of restarting the diagnosis in critical cases. To assist clinicians in selecting high-quality ultrasound images and ensuring accurate diagnoses, we introduce Ultrasound-QBench, a comprehensive benchmark that systematically evaluates multimodal large language models (MLLMs) on quality assessment tasks of ultrasound images. Ultrasound-QBench establishes two datasets collected from diverse sources: IVUSQA, consisting of 7,709 images, and CardiacUltraQA, containing 3,863 images. These images encompassing common ultrasound imaging artifacts are annotated by professional ultrasound experts and classified into three quality levels: high, medium, and low. To better evaluate MLLMs, we decompose the quality assessment task into three dimensionalities: qualitative classification, quantitative scoring, and comparative assessment. The evaluation of 7 open-source MLLMs as well as 1 proprietary MLLMs demonstrates that MLLMs possess preliminary capabilities for low-level visual tasks in ultrasound image quality classification. We hope this benchmark will inspire the research community to delve deeper into uncovering and enhancing the untapped potential of MLLMs for medical imaging tasks.
\end{abstract}


\section{Introduction}
\label{sec:intro}

Ultrasound imaging represents a medical imaging technology that is prevalently utilized in clinical practice. The advantages of ultrasound imaging such as low cost, ease of operation, radiation-free nature, and the ability to provide real-time imaging, render it ideal for visualizing soft tissues \cite{wu2017fuiqa}. Currently, ultrasound imaging has been extensively employed for diagnosing abdominal, cardiac, vascular, and musculoskeletal diseases, as well as for prenatal examinations. Nevertheless, with the increasing volume of daily ultrasound examinations, the variability in image quality has emerged as a considerable challenge that affects diagnostic accuracy, data management, and healthcare efficiency \cite{nikolaev2021quantitative}. Low-quality images not only diminish diagnostic precision but also lead to repeat scans, thereby increasing healthcare costs and resource wastage \cite{song2022medical}.

In order to evaluate image quality, quality assessment (QA) has been extensively investigated in the domain of natural images. In traditional quality assessment methods, image features are manually extracted and statistically analyzed, and subsequently compared with reference images for full-reference methods such as Structural Similarity Index (SSIM) \cite{wang2004image} or directly estimated for no-reference methods such as BRISQUE \cite{mittal2012no} and NIQE \cite{mittal2012making}. In contrast to natural images, quality assessment of medical images involves diagnostic relevance and the handling of artifacts specific to medical images, such as noise and blurring \cite{rodrigues2024objective}, rendering traditional methods less effective. Specifically, ultrasound images often incorporate distinctive artifacts resulting from operator variability and diverse imaging conditions \cite{chen2021muiqa, ohashi2023applicability} such as multiple reflections, multiple internal reverberations, and refractive shadow. Consequently, merely reutilizing the image features employed in natural image quality assessment without taking into account the specialized characteristics of ultrasound imaging is unable to address these challenges \cite{dong2019generic}.
To further enhance the generalization of traditional methods, researchers explore the capabilities of neural networks in quality assessment. These methods convert quality assessment tasks into an end-to-end classification or regression problem, and substitute traditional hand-crafted feature extraction with learning-based feature representations. However, the learning process relies on a considerable amount of labeled data, which is costly and scarce in the context of medical imaging, and the assessment performance may deteriorate in the presence of unlabeled noise and artifacts. 

\begin{figure*}[h]
  \centering
  \includegraphics[width=0.95\textwidth]{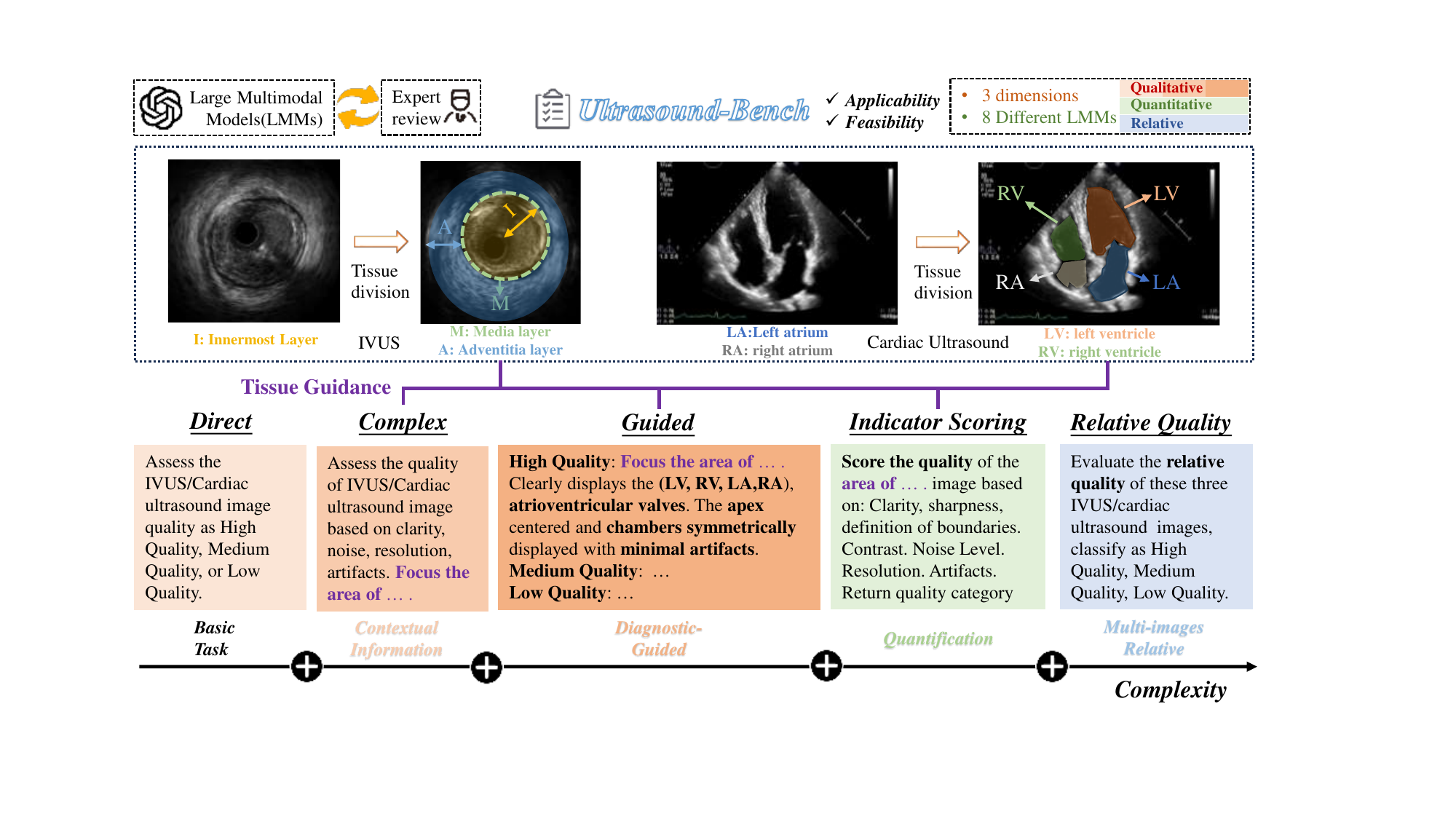}
  \caption{In the proposed Ultrasound-QBench, we established the first benchmark on MLLM capabilities
On ultrasound images, qualitative quality assessment, quantitative evaluation, and relative quality are included.}
  \label{fig:framework}
\end{figure*}

Based on the aforementioned analysis, an ideal quality assessment method of ultrasound images requires to be both specialized and generalizable. The former necessitates that the quality assessment method fully utilize the characteristics of ultrasound imaging, while the latter requires that the quality assessment method be capable of making accurate judgments for different types of artifacts. Recent progress in Multimodal Large Language Models (MLLMs), such as LLaVA \cite{liu2024visual}, Qwen2-vl \cite{wang2024qwen2}, and mPLUG-owl3 \cite{ye2024mplug}, holds great promise in medical image quality assessment due to their zero-shot inference capacity and cross-domain expertise. Unlike traditional neural networks that require a large amount of labeled data for training, MLLMs can leverage generative pre-trained models and given prompts to conduct inference, demonstrating excellent performance with limited labeled data or even in zero-shot scenarios \cite{li2024llava, saab2024capabilities,tang2023medagents, chen2023towards}. This suggests the potential for MLLMs to accurately assess ultrasound image quality, even in the presence of unobserved distortion types. Moreover, the pre-trained models have learned from an extensive amount of labeled data across diverse fields, presenting exceptional cross-domain expertise. Based on these advantages, MLLMs are both specialized in cross-domains and generalizable to unencountered samples, making them particularly appropriate for ultrasound image quality assessment.

Inspired by the quality evaluation experiments on natural images using MLLMs\cite{Wu2024QBench, Wu2024QInstruct, Wu2024QAlign}, we propose a new benchmark framework, \textbf{Ultrasound-QBench}, to evaluate MLLMs in ultrasound image quality assessment. In summary, Ultrasound-QBench investigates the generalization capacity of MLLMs to ultrasound image quality assessment without task-specific fine-tuning, bridging the gap between natural image QA and medical QA. Fig.~\ref{fig:framework} present the overall diagram of Ultrasound-QBench framework. We evaluate 7 open-source MLLMs as well as 1 proprietary MLLM from three dimensionalities:

\begin{itemize}
    \item \textbf{Qualitative Classification:}  
    MLLMs are required to classify the ultrasound images into three quality degrees: low, medium, and high. Prompts spanning from rough to diagnostic level are utilized to assist MLLMs in making judgments. As depicted in Fig.~\ref{fig:framework}, the prompts are categorized into three complexity levels: basic question, contextual information, and diagnostic-guided information. For the second and third tasks, the tissues information is included in the prompts.

    \item \textbf{Quantitative Scoring:}  
    MLLMs are required to provide quantitative scores in multiple quality-related indicators such as clarity, contrast, noise level, detail resolution, uniformity, and presence of artifact. As demonstrated in Fig.~\ref{fig:framework}, the prompts contain the tissues information, guiding the MLLMs to focus on the region of interest.
    
    \item \textbf{Comparative Assessment:}  
    MLLMs are required to conduct a comparison of the relative quality among multiple ultrasound images. This task involves evaluating the ability of MLLMs to comprehend the relative quality changes between diverse images.
\end{itemize}

Ultrasound-QBench establishes two datasets collected from diverse sources: IVUSQA, comprising 7,709 images, and CardiacUltraQA, containing 3,863 images. These images encompass common ultrasound imaging artifacts and noises, e.g. multiple reflections, multiple internal reverberations, and refractive shadow, and are annotated by professional ultrasound experts and classified into three quality levels: high, medium, and low. By addressing the limitations of traditional quality assessment methods and leveraging the strengths of MLLMs, this work lays the foundation for advancing automated quality assessment of ultrasound images. Our findings provide valuable insights into the potential of MLLMs for optimizing diagnostic workflows in clinical practice.

\section{Dataset}

\subsection{Overview}

To evaluate the performance of multimodal large language models (MLLMs) in ultrasound image quality assessment, we establish two ultrasound image datasets that represent two typical real-world clinical scenarios: IVUSQA, concentrating on intravascular ultrasound (IVUS) images, and CardiacUltraQA, covering cardiac ultrasound images. These two datasets are curated to represent diverse imaging conditions and feature high-quality expert annotations, providing a robust framework for zero-shot testing.

\begin{figure}[H]
  \centering
  \includegraphics[width=0.45\textwidth]{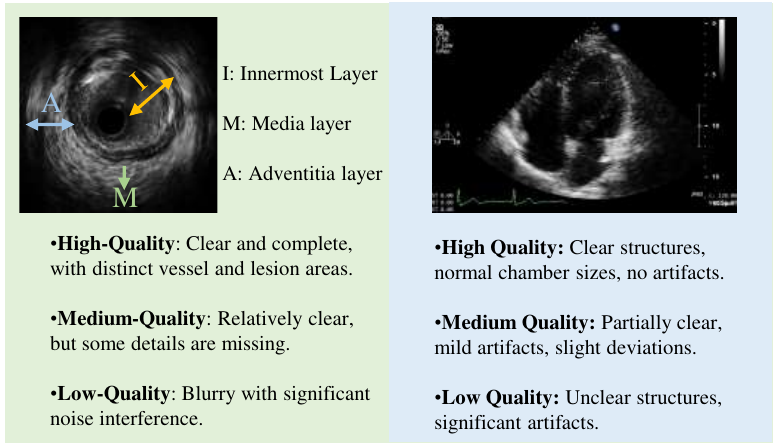}
  \caption{IVUSQA and CardiacUltraQA Dataset and Assessment Standard.}
  \label{fig:Assessment Standard}
\end{figure}

\subsection{Dataset Composition}

\textbf{IVUSQA Dataset:}  
This dataset contains a total of 7,709 intravascular ultrasound (IVUS) images, which are used to assess vascular structures. IVUS imaging is characterized by a circular view, centered on the vessel wall, and is mainly used to detect plaques and other abnormalities in arteries. The three primary components of an IVUS image include: \textbf{(1) Innermost Layer:} this layer comprises the intima, atheroma, and internal elastic membrane; \textbf{(2) Media Layer:} composed of smooth muscle cells that do not reflect ultrasound, appearing as dark areas in the image; \textbf{(3) Adventitia Layer:} this layer consists of collagen, which reflects a significant amount of ultrasound, presenting as white in IVUS images.

\textbf{CardiacUltraQA Dataset:}  
The CardiacUltraQA dataset consists of 3,863 the Apical Four-Chamber View images. This view displays the four main chambers of the heart (left atrium, right atrium, left ventricle, right ventricle) along with associated structures such as the heart valves and surrounding tissues. These images are crucial for assessing heart health, particularly for observing the symmetry of the heart chambers and normal anatomical features.

\begin{figure}[t]
  \centering
  \includegraphics[width=0.45\textwidth]{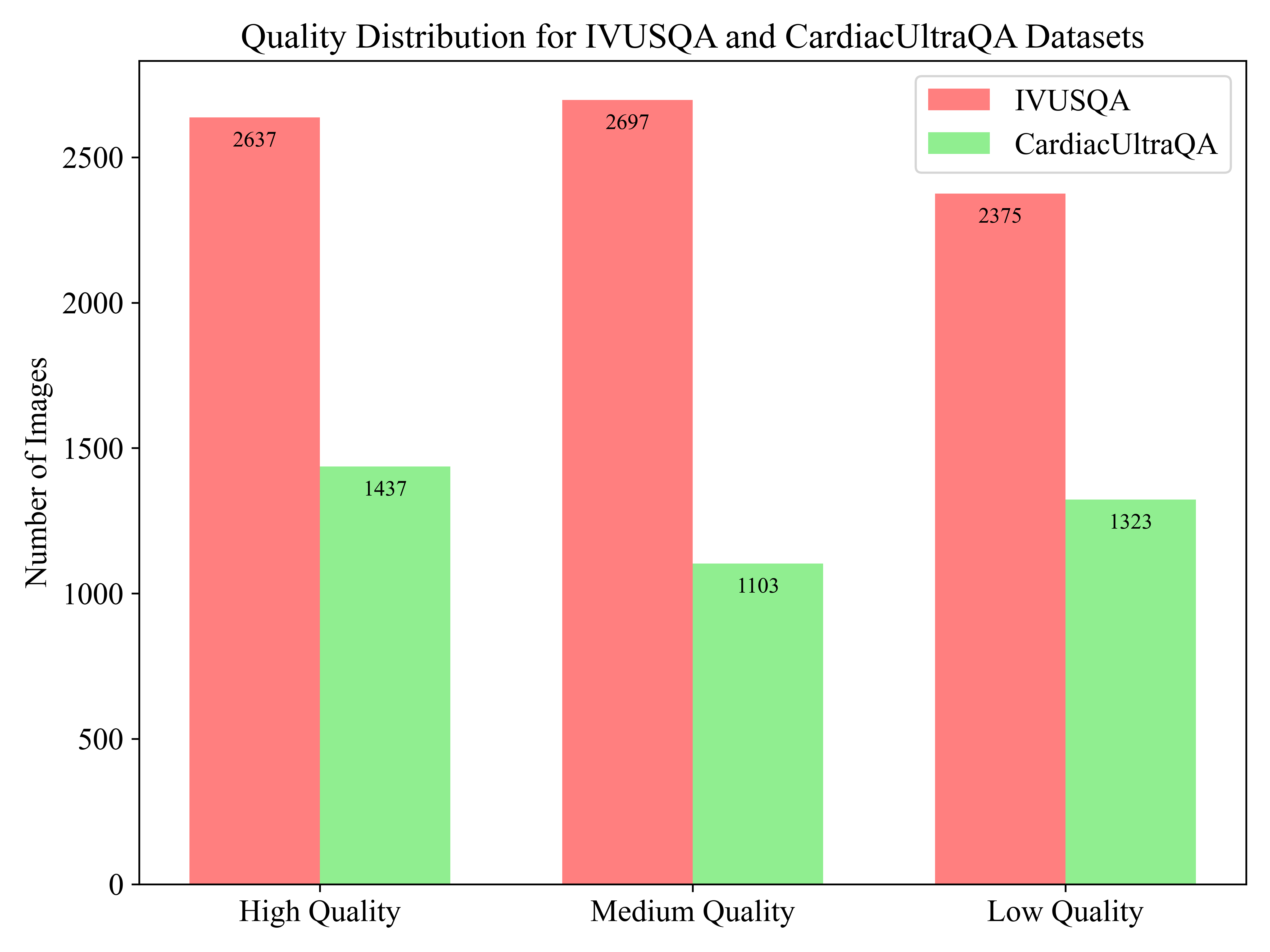}
  \caption{Quality Distribution combined for IVUSQA and CardiacUltraQA Dataset.}
  \label{fig:dataset_distribution}
\end{figure}

The quality of images in these two datasets is classified into three categories: high, medium, and low, based on the standard illustrated in Fig.~\ref{fig:Assessment Standard}. All images are annotated by a team of certified ultrasound experts using a unified and standardized subjective quality assessment framework to ensure consistency and clinical relevance. Fig.~\ref{fig:dataset_distribution} depicts the subjective quality distribution of these two datasets.

\subsection{Advantages of the Dataset}

\begin{itemize}
    \item\textbf{Clinical Significance:}  
   Both IVUSQA and CardiacUltraQA focus on critical diagnostic areas, including vascular health and cardiac function, ensuring the applicability of the evaluation results in clinical practice.
   
    \item\textbf{Data Diversity:}  
   The images in the dataset originate from various operators and imaging conditions, thereby incorporating a wide range of noise and artifacts commonly encountered in ultrasound imaging.
   
    \item\textbf{Expert Annotations:}  
   The quality label attached to each image is obtained by multiple ultrasound physicians and medical doctoral students according to a standardized and unified subjective assessment process, ensuring high reliability in clinical practice.

\end{itemize}

\section{Evaluation Workflow}

\subsection{Prompt Design}

To systematically evaluate the performance of Multimodal Large Language Models (MLLMs) in ultrasound image quality assessment, we decompose the evaluating task into three sub-tasks: (1) qualitative classification, (2) quantitative scoring, and (3) comparative assessment.

\begin{figure*}[h]
  \centering
  \includegraphics[width=0.95\textwidth]{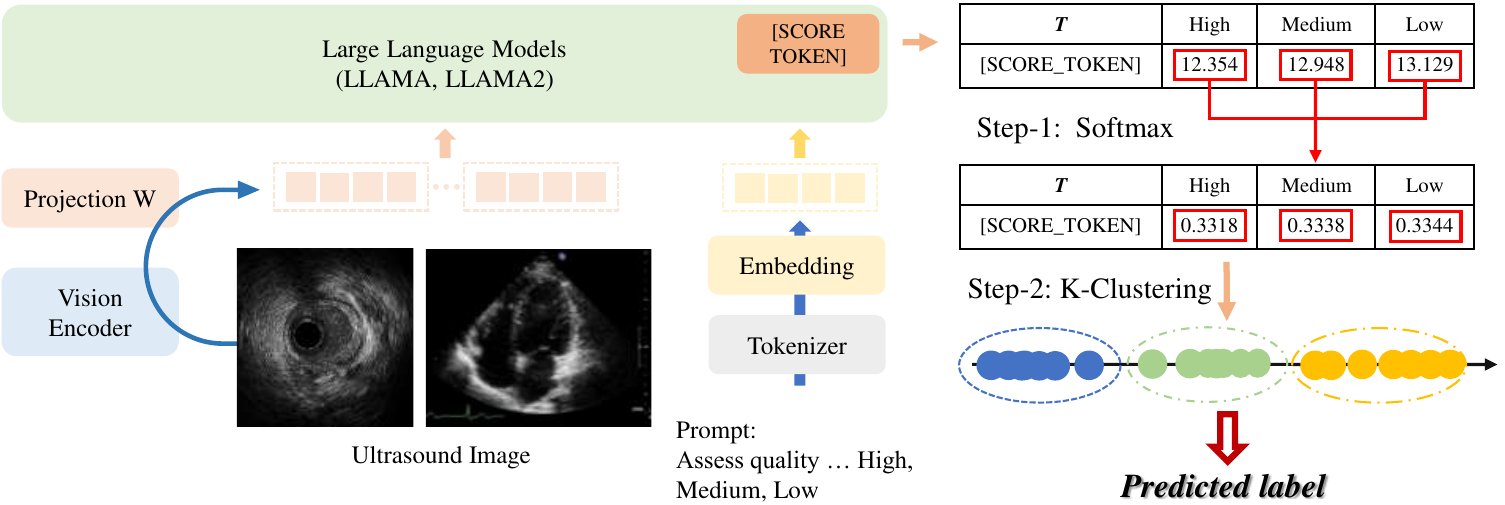}
  \caption{The proposed softmax-based quality assessment strategy for MLLMs improves upon existing methods by extracting logits for the 'high quality,' 'medium quality,' and 'low quality' categories, rather than directly decoding tokens from the [SCORE TOKEN] position. The strategy predicts labels through a weighted summation and pooling of these logits, followed by a weighted clustering to obtain the final quality rating.}
  \label{fig:IVUSQA_softmax_clustering}
\end{figure*}

\paragraph{\textbf{Qualitative Classification}} In this sub-task, MLLMs are required to classify ultrasound images into three quality categories: low, medium, and high. As illustrated in the three \textbf{\textcolor[rgb]{0.956, 0.694, 0.513}{orange}} boxes of Fig.~\ref{fig:framework}, prompts of varying complexity are provided to assist MLLMs in assessing image quality: basic questions, contextual details, and diagnostic guidance. For the second and third tasks, additional tissue information is integrated into the prompts to help MLLMs focus on critical regions. These three types of prompts are described in detail as follows:

\begin{itemize}
    \item \textbf{basic questions:} the model's baseline performance in image quality assessment without additional context or guidance is evaluated. 
    
    \item \textbf{contextual details:} additional contextual information, such as the anatomical region of the ultrasound image or the clinical use case, is provided to the model. The impact of more detailed contextual descriptions on the model's ability to enhance classification accuracy and its understanding of task-specific information is evaluated.
    
    \item \textbf{diagnostic guidance:} specific diagnostic criteria is introduced to guide the classification process. 
    
    These prompts evaluate the model's ability to follow detailed clinical instructions and make classification decisions based on predefined diagnostic criteria.

\end{itemize}

\paragraph{\textbf{Quantitative Scoring}} This sub-task requires the model to provide an overall quality score based on multiple quality indicators such as clarity, noise level, resolution, thereby evaluating the model's ability to perform multi-dimensional reasoning. The \textbf{\textcolor[rgb]{0.663, 0.819, 0.557}{green}} box of Fig.~\ref{fig:framework} presents a typical template of this kind of prompt.
    
\paragraph{\textbf{Comparative Assessment}} This sub-task challenges the model to compare multiple images and identify their relative quality differences. The \textbf{\textcolor[rgb]{0.561, 0.667, 0.863}{blue}} box of Fig.~\ref{fig:framework} presents a typical template of this kind of prompt.  

\begin{table*}
\centering
\caption{Accuracy for Ultrasound Image Quality Assessment Across Two Datasets with Five Prompts. \textcolor{red}{Red} indicates the best result in each task, and \underline{underlined} values represent the most significant improvement with Softmax + Clustering.}
\scalebox{0.72}{
\begin{tblr}{
  row{1} = {c},
  row{2} = {c},
  row{3} = {c},
  cell{1}{1} = {r=3}{},
  cell{1}{2} = {c=5}{},
  cell{1}{7} = {c=5}{},
  cell{2}{2} = {c=3}{},
  cell{2}{5} = {r=2}{},
  cell{2}{6} = {r=2}{},
  cell{2}{7} = {c=3}{},
  cell{2}{10} = {r=2}{},
  cell{2}{11} = {r=2}{},
  cell{4}{1} = {l},
  cell{4}{2} = {c},
  cell{4}{3} = {c},
  cell{4}{4} = {c},
  cell{4}{5} = {c},
  cell{4}{6} = {c},
  cell{4}{7} = {c},
  cell{4}{8} = {c},
  cell{4}{9} = {c},
  cell{4}{10} = {c},
  cell{4}{11} = {c},
  cell{5}{2} = {c},
  cell{5}{3} = {c},
  cell{5}{4} = {c},
  cell{5}{5} = {c},
  cell{5}{6} = {c},
  cell{5}{7} = {c},
  cell{5}{8} = {c},
  cell{5}{9} = {c},
  cell{5}{10} = {c},
  cell{5}{11} = {c},
  cell{6}{2} = {c},
  cell{6}{3} = {c},
  cell{6}{4} = {c},
  cell{6}{5} = {c},
  cell{6}{6} = {c},
  cell{6}{7} = {c},
  cell{6}{8} = {c},
  cell{6}{9} = {c},
  cell{6}{10} = {c},
  cell{6}{11} = {c},
  cell{7}{2} = {c},
  cell{7}{3} = {c},
  cell{7}{4} = {c},
  cell{7}{5} = {c},
  cell{7}{6} = {c},
  cell{7}{7} = {c},
  cell{7}{8} = {c},
  cell{7}{9} = {c},
  cell{7}{10} = {c},
  cell{7}{11} = {c},
  cell{8}{2} = {c},
  cell{8}{3} = {c},
  cell{8}{4} = {c},
  cell{8}{5} = {c},
  cell{8}{6} = {c},
  cell{8}{7} = {c},
  cell{8}{8} = {c},
  cell{8}{9} = {c},
  cell{8}{10} = {c},
  cell{8}{11} = {c},
  cell{9}{2} = {c},
  cell{9}{3} = {c},
  cell{9}{4} = {c},
  cell{9}{5} = {c},
  cell{9}{6} = {c},
  cell{9}{7} = {c},
  cell{9}{8} = {c},
  cell{9}{9} = {c},
  cell{9}{10} = {c},
  cell{9}{11} = {c},
  cell{10}{2} = {c},
  cell{10}{3} = {c},
  cell{10}{4} = {c},
  cell{10}{5} = {c},
  cell{10}{6} = {c},
  cell{10}{7} = {c},
  cell{10}{8} = {c},
  cell{10}{9} = {c},
  cell{10}{10} = {c},
  cell{10}{11} = {c},
  cell{11}{2} = {c},
  cell{11}{3} = {c},
  cell{11}{4} = {c},
  cell{11}{5} = {c},
  cell{11}{6} = {c},
  cell{11}{7} = {c},
  cell{11}{8} = {c},
  cell{11}{9} = {c},
  cell{11}{10} = {c},
  cell{11}{11} = {c},
  vline{2} = {1}{},
  vline{2,3,5-8,10,11} = {2}{},
  vline{2,5,6,7,10-11} = {3}{},
  vline{2,5-7,10-11} = {4-11}{},
  hline{1,4} = {1-11}{},
  hline{2} = {2-11}{},
  hline{3} = {2-4,7-9}{},
  hline{11} = {-}{dashed},
  hline{12} = {-}{},
}
MLLM         & IVUSQA      &            &            &               &              & CardiacUltraQA &            &            &               &              \\
                             & Qualitative &            &            & Quantitative~ & Comparative~ & Qualitative    &            &            & Quantitative~ & Comparative~ \\
                             & Basic       & Contextual & Diagnostic &               &              & Basic          & Contextual & Diagnostic &               &              \\
LLaVA-v1.5-7B   & \textcolor{red}{\textbf{34.27\%}}/46.52\%     & 34.23\%/46.63\%    & 34.21\%/40.36\%    & 34.31\%/\underline{48.71\%}       & 94.06\%      & 37.21\%/44.74\%        & 37.32\%/\underline{51.51\%}    & 37.26\%/\textcolor{red}{\textbf{48.44\%}}    & 37.35\%/\textcolor{red}{\underline{\textbf{50.14\%}}}
       & 97.28\%      \\
LLaVA-v1.5-13B  & 27.14\%/47.65\%     & 34.99\%/\textcolor{red}{\underline{\textbf{50.11\%}}}    & 35.06\%45.88\%    & \textcolor{red}{\textbf{34.99\%}}/\textcolor{red}{\textbf{49.25\%}}       & 97.60\%      & \textcolor{red}{\textbf{40.31\%}}/46.27\%        & 31.14\%/45.01\%    & 35.14\%/42.58\%    & \textcolor{red}{\textbf{40.37\%}}/46.33\%       & 98.45\%      \\
InternLM-VL        & 34.20\%/41.47\%     & 30.81\%/41.99\%    & 34.21\%/40.74\%    & 34.35\%/42.86\%       & 55.64\%      & 36.86\%/44.07\%        & 37.67\%/45.57\%    & 37.07\%/45.46\%    & 38.15\%/44.29\%       & 80.27\%      \\
Deepseek                     & 32.46\%/42.38\%     & 34.27\%/43.57\%    & 32.53\%/41.35\%    & 34.21\%/46.37\%       & 49.64\%      & 39.71\%/42.58\%        & 37.19\%/43.93\%    & 37.21\%/46.74\%    & 37.20\%45.52\%       & 73.54\%      \\
Qwen2-VL             & 30.90\%/41.19\%     & \textcolor{red}{\textbf{36.99\%}}/42.63\%    & 30.82\%/\underline{45.65\%}    & \textcolor{red}{\textbf{34.99\%}}/45.93\%       & 98.94\%      & 38.99\%/\textcolor{red}{\underline{\textbf{50.01\%}}}
        & \textcolor{red}{\textbf{55.94\%}}/\textcolor{red}{\textbf{61.37\%}}    & 34.25\%/\underline{45.90\%}    & 37.21\%/45.18\%       & 99.35\%      \\
mPLUG-Owl3        & 31.08\%/\textcolor{red}{\underline{\textbf{51.38\%}}}
     & 34.17\%/36.53\%    & \textcolor{red}{\textbf{57.88\%}}/\textcolor{red}{\textbf{64.04\%}}    &   34.21\%/47.43\%            & 98.66\%      & 38.67\%/48.49\%        & 38.78\%/46.37\%    & \textcolor{red}{\textbf{44.63\%}}/45.45\%    &      38.67\%/47.99\%         & 99.64\%      \\
LLaVA-Med        & 34.21\%/46.26\%     & 34.21\%/39.64\%    & 34.99\%/40.03\%    &    34.21\%/41.74\%           & 96.42\%      & 38.67\%/46.26\%        & 38.78\%/42.63\%    & 37.21\%/43.16\%    &     37.21\%/44.61\%          & 99.88\%      \\
GPT-4o (proprietary)         & 35.63\%     & 38.41\%    & 41.24\%    & 41.31\%       & 99.93\%      & 40.67          & 42.13\%    & 45.37\%    & 44.28\%       & 99.96\%      
\end{tblr}}
\label{tab:performance}
\end{table*}

\subsection{Experimental Workflow}

The experimental workflow is organized into three stages: model selection, execution, and evaluation.

\paragraph{\textbf{Model Selection Stage}} In this stage, we select 7 open-source MLLMs as well as 1 proprietary MLLM to evaluate their capacities of quality assessment for IVUSQA and CardiacUltraQA. These models include LLaVA-v1.5-7b, LLaVA-v1.5-13b, InternLM-XComposer2-VL-7b\cite{Dong2024}, DeepSeek\cite{liu2024deepseek}, LLaVA-Med, Qwen2-VL, mPLUG-Owl3, and GPT4o (proprietary). The five kinds of prompts corresponding to the three sub-tasks are predefined to ensure consistency across the experiments. Optimal settings for each model are selected to ensure reliable inference results.

\paragraph{\textbf{Execution Stage}} In this stage, ultrasound images are processed in batches of 16 on an NVIDIA RTX 4090 GPU with 24GB of memory. For each image-prompt pair, the evaluated model generates a response, which is then parsed to extract the score tokens. To improve the assessment accuracy and prevent predictions from being biased toward extreme outcomes, we propose a new evaluation strategy that combines Softmax and k-Clustering, as illustrated in Fig.~\ref{fig:IVUSQA_softmax_clustering}. The strategy consists of two steps:

\begin{itemize}
    \item \textbf{Step-1 Softmax Strategy:} The model outputs score\_tokens for each quality level based on the prompt. By applying the Softmax operation, the raw logits are transformed into comparable probability values, reducing the impact of extreme values on the model's predictions and preventing the model from excessively favoring certain classes.
    \item \textbf{Step-2 Clustering:} The clustering method is combined with a weighted summation approach to optimize the predictions output by Step-1. By grouping similar prediction results together, clustering effectively prevents the model from excessively favoring any one class, ensuring balanced prediction results. Clustering can automatically adjust the class boundaries based on the distribution of the prediction results, reducing the risk of model bias due to extreme values. This helps to ensure a more balanced distribution of samples across each class and improves the accuracy and stability of the model's classification.

\end{itemize}

\paragraph{\textbf{Evaluation Stage}} The evaluation stage computes the key metric of classification accuracy and assesses prompt sensitivity to evaluate the models' performance across different strategies. The analysis includes a comparison of model performance on the two datasets to assess the models' generalizability across different types of ultrasound images. Prompt sensitivity is evaluated by examining how the models' performance varies with different prompt strategies.

\section{Experimental results}
This section presents and analyze the performance of selected 8 MLLMs on ultrasound image quality assessment. The evaluating results of the selected models on three tasks are presented in Table~\ref{tab:performance}.

\subsection{Original Evaluation without Softmax and Clustering}
We first analyze the original performance of MLLMs without Softmax and k-Clustering.
\paragraph{\textbf{Qualitative Classification}}
As shown in Table~\ref{tab:performance}, the original capabilities of the selected MLLMs to qualitatively assess ultrasound image quality are disappointing. For basic questions, all of these models exhibit a poor accuracy of around 30\% on both IVUSQA and CardiacUltraQA. Through analyzing these results, we find an obvious bias toward a specific class. For instance, mPLUG-Owl tends to classify all images as high quality. For complex contextual information, Qwen2-VL achieves 55.94\% accuracy on CardiacUltraQA, while other models still perform poorly. After introducing the diagnostic-level standard into prompts, only mPLUG-Owl3 can improve its accuracy to 57.88\%, with the accuracy of medium-quality and low-quality images reaching approximately 50\%. These results indicate that existing MLLMs struggle to understand the classification standard of ultrasound image quality in the same way as humans.

\paragraph{\textbf{Quantitative Scoring}} 
As shown in Table~\ref{tab:performance}, all of these models perform poorly in predicting accurate quality scores, indicating that existing MLLMs struggle to understand the definitions of quality-related indicators.

\paragraph{\textbf{Comparative Assessment}}
As shown in Table~\ref{tab:performance}, all of these models can accurately distinguish the relative quality of the given image sequences, indicating that these models can detect changes in ultrasound image quality.

\subsection{Evaluation with Softmax and Clustering}
We performed an extensive evaluation of the performance of Multimodal Large Language Models (MLLMs) augmented with a Softmax strategy enhanced by k-means clustering, with the experimental results presented in Table~\ref{tab:performance}. Prior to the incorporation of this strategy, the output distributions across tasks exhibited significant imbalance. For instance, LLaVA-v1.5-7B tended to predict high-quality classifications predominantly across all tasks, except for the relative evaluation task. Upon integrating k-means clustering with the Softmax approach, the model outputs became more balanced, effectively mitigating the tendency of the model to excessively rely on high-quality predictions. This strategy resulted in an average improvement of approximately 10.11\% in accuracy across all evaluation prompts. In particular, mPLUG-Owl3 demonstrated a notable accuracy gain of 20.3\% in the basic task. The proposed approach successfully mitigates the adverse effects of imbalanced output distributions, thereby enhancing the overall balance of predictions and increasing the robustness of the model’s performance. In Table I, the best performance for each task is marked in \textcolor{red}{\textbf{red}}, while the most significant improvements attributed to this strategy are \underline{\textbf{underlined}} for each task.

\section{Discussion}

\subsection{Limitations}

Despite the improvements achieved with Softmax + Clustering, the performance of MLLMs in quantitative and qualitative assessment of ultrasound image quality remains unsatisfactory. For qualitative assessment, these models still exhibit limitations in assessment accuracy, particularly in distinguishing between medium and low-quality images. For instance, models such as mPLUG-Owl have a tendency to overclassify images as high quality, which indicates an inadequate ability to accurately assess different quality levels. In terms of quantitative scoring, all models fail to predict accurate quality scores reliably. This highlights the necessity for enhanced domain-specific knowledge and advanced feature extraction techniques to improve overall performance. Furthermore, we believe there are two additional factors that limit the performance of MLLMs on the task of ultrasound image quality assessment:

\paragraph{Insufficient Understanding of Ultrasound Image Features}  
Despite prompt guidance, current MLLMs lack a full understanding of ultrasound-specific features such as speckle noise, operator variability, and artifacts. These features complicate accurate classification, particularly for high-quality images. Fine-tuning with high-quality, domain-specific data is necessary for models to better capture these unique characteristics.


\paragraph{Dependence on Prompt Engineering}  
MLLMs heavily rely on carefully designed prompts, which limits their adaptability in real-world scenarios where input quality can vary. Ultrasound images exhibit significant differences in resolution, contrast, and artifacts. This reliance reduces model flexibility, causing performance degradation when faced with simpler or less structured prompts. Enhancing model adaptability is crucial for real-world medical image assessments.




\subsection{Future Research Directions}

Future research should focus on the following areas:

\paragraph{Reducing Dependence on Prompt Engineering}  
Developing adaptive methods to reduce reliance on specific prompts can enhance model flexibility and applicability in real-world medical scenarios.

\paragraph{Expanding Dataset Diversity}  
In order to improve model generalization and performance on low-quality images, we will further expand ultrasound datasets to include diverse imaging modalities, patient groups, and clinical conditions.

\paragraph{Leveraging Domain-Specific Knowledge}
By fine-tuning with high-quality labeled data, we can incorporate ultrasound-specific knowledge into model training. Additionally, exploring techniques such as few-shot learning and self-supervised learning can improve performance in data-limited scenarios, thereby enhancing both qualitative and quantitative assessments.

\section{Conclusion}

This paper presents Ultrasound-QBench, a benchmark for evaluating multimodal large language models (MLLMs) in ultrasound image quality assessment using the established IVUSQA and CardiacUltraQA datasets. Eight MLLMs are evaluated across qualitative classification, quantitative scoring, and comparative assessment tasks. While models like mPLUG-Owl and Qwen2 show potential in judging ultrasound image quality like humans, they struggle with accurately distinguishing between image quality levels and comprehending ultrasound-specific features, such as distortions and noise. The Softmax + Clustering method can improve accuracy by 10.11\%, but limitations remain in understanding ultrasound image structure.

Future work should focus on reducing dependency on prompt engineering, enhancing dataset diversity to improve generalization, and incorporating domain-specific knowledge. Fine-tuning with high-quality labeled data, along with techniques like few-shot learning and self-supervised learning, will further strengthen MLLMs' performance in ultrasound image quality assessment.

\bibliographystyle{IEEEbib}
\bibliography{Ultrasound-QBench}

\vspace{12pt}
\color{red}

\end{document}